\documentclass[11pt,a4paper]{article}
\usepackage{amsmath}
\usepackage{amsfonts,amssymb}
\begin{document}

\newcommand{\nl}{\nonumber\\}
\newcommand{\bes}{\begin{eqnarray}}
\newcommand{\eens}{\nonumber\end{eqnarray}}

\renewcommand{\/}{\over}
\renewcommand{\d}{\partial}
\newcommand{\im}{\hbox{im\,}}

\newcommand{\eps}{\epsilon}
\newcommand{\vth}{\vartheta}
\renewcommand{\th}{\theta}
\newcommand{\dlt}{\delta}
\newcommand{\Om}{\Omega}

\newcommand{\dmu}{\d_\mu}
\newcommand{\dnu}{\d_\nu}
\newcommand{\xmu}{\xi^\mu}

\newcommand{\four}[1]{{\hbox{$4{#1}\/3$}}}

\newcommand{\dd}[1]{{\d\/\d #1}}

\newcommand{\toj}{\tilde\oj}
\newcommand{\oj}{{\mathfrak g}}
\newcommand{\tr}{\hbox{tr}\,}

\newcommand{\ssu}{\allowbreak su(3)+su(2)+u(1)}
\newcommand{\vect}{{\mathfrak{vect}}}
\newcommand{\map}{{\mathfrak{map}}}
\newcommand{\mb}{{\mathfrak{mb}}}
\newcommand{\ksle}{{\mathfrak{ksle}}}
\newcommand{\vle}{{\mathfrak{vle}}}
\newcommand{\kk}{{\mathfrak{k}}}
\newcommand{\kas}{{\mathfrak{kas}}}

\newcommand{\barr}{\begin{array}}
\newcommand{\earr}{\end{array}}

\newcommand{\NN}{{\mathbb N}}
\newcommand{\CC}{{\mathbb C}}
\newcommand{\RR}{{\mathbb R}}
\newcommand{\ZZ}{{\mathbb Z}}

\title{{Maximal depth implies su(3)+su(2)+u(1)}}

\author{T. A. Larsson \\
Vanadisv\"agen 29, S-113 23 Stockholm, Sweden,\\
 email: Thomas.Larsson@hdd.se }

\maketitle 
\begin{abstract} 
Hence it excludes proton decay and supersymmetry. The main
predictions of a gauge model based on the exceptional simple Lie
superalgebra $\mb(3|8)$ (a localized version of $\ssu$) are reviewed.
\end{abstract}



\newpage

\renewcommand{\arraystretch}{1.4}

Simple infinite-dimensional Lie superalgebras of polynomial vector fields
have recently been classified \cite{Kac98,Sh99}. The exception $\mb(3|8)$
(a.k.a. $E(3|8)$) \cite{CK99,Lar01,Sh83} is singled out as the unique
algebra of maximal depth $3$ in its consistent $\ZZ$-gradation. Its
representation theory was recently worked out \cite{GLS01,KR01,Lar02}.
This superalgebra, and its close relative $\vle(3|6) = E(3|6)$ of depth
$2$, are Cartan prolongs $(\oj_-, \oj_0)_*$ where $\oj_-$ is nilpotent
and $\oj_0 = \ssu$\footnote{Technically, the classification concerns superalgebras 
over the field of complex numbers, so $\oj_0$ is really 
$sl(3)+sl(2)+gl(1)$, the non-compact form of $\ssu$. The problem of
choosing the right real form of $\mb(3|8)$ is not considered here.}.
This means that there is a 1-1 correspondence between $\mb(3|8)$ (and
$\vle(3|6)$) irreps and $\ssu$ irreps, suggesting that an $\mb(3|8)$
symmetry might be mistaken experimentally for the symmetries of the
standard model (SM) in particle physics.

With this in mind, I recently constructed a gauge model with $\mb(3|8)$
symmetry \cite{Lar02}. It was called the ``second-gauged standard model''
(SGSM), since the $\ssu$ symmetry is made local not only in spacetime
(first gauging), but also in internal space (second gauging). However,
the details of this model are not necessary to extract the main
experimental consequences, which differ sharply from predictions of
popular theories such as supersymmetry (SUSY) and grand unified theories
(GUTs). It is the purpose of this note to list and motivate these
predictions, which will be tested at the LHC and at SuperKamiokande over
the next decade. My hope is also that model builders will be attracted by
this substantially new {\em local} mathematics, which immediately and
unambigously leads to the correct SM symmetries. 
Technical details are given in the appendix.

The main observation is that the correct symmetry of the SM is not $\ssu$
itself, but rather its associated current algebra $\map(4, \ssu)$ of maps
from four-dimensional spacetime to the gauge algebra. Since the bosons in
the SM are gauge bosons, i.e. current algebra connections, the bosons in
the SGSM must be connections of $\map(4,\mb(3|8))$; this point was missed
in \cite{Kac99}. Just as there is a 1-1 correspondence between $\mb(3|8)$
and $\ssu$ irreps, there is also a 1-1 correspondence between connections
in the associated current algebras.

The SGSM thus predicts absense of new gauge bosons, and consequently no
proton decay, except possibly by instanton effects. In contrast, all
GUTs, and theories which can be approximated by GUTs at low energies,
have interactions that do not conserve baryon number; in fact, the
minimal $SU(5)$ model was slayed long ago precisely because it predicted
too rapid proton decay. Also low-energy SUSY models, both the MSSM and
the ESSM, predict proton decay at rates which are straining the bounds
from SuperKamiokande \cite{Pati02}. Within a decade, proton decay will
either be discovered, or the bounds will be stringent enough to rule out
GUTs and SUSY.

The main reason why SUSY became popular is that it is the only (known)
way to circumvent the Coleman-Mandula theorem and combine internal and
spacetime symmetries in a non-trivial way \cite{West90}. However, there
is no good experimental reason why such unification would be desirable.
On the contrary, if we elevate maximal depth to a guiding principle (and
at least the $\ssu$ part is strongly supported by experiments),
internal/spacetime unification is ruled out. Although there certainly are
simple Lie superalgebras containing both $\mb(3|8)$ and the Poincar\'e
algebra as subalgebras ($\vect(7|8)$ is an obvious example), any
such superalgebra is necessarily of less than maximal depth. Hence
maximal depth requires that the $\map(4,\mb(3|8))$ symmetry be combined
with spacetime symmetries in the trivial way, i.e. as a semi-direct
product. GUTs are excluded for essentially the same reason. Any
finite-dimensional Lie algebra $\oj \supset \ssu$ is of less depth, as
are all prolongs based on $\oj$.

If internal and spacetime symmetries are not combined non-trivially,
gravity is not unified with other interactions. Since the main argument
for string theory is that it allegedly contains gravity \cite{Pol98},
maximal depth and string theory are mutually exclusive. $\mb(3|8)$ is
proposed as a generalization of the SM symmetries, whereas gravity must
be dealt with separately. It can be remarked that all projective
lowest-energy Fock representations of the spacetime diffeomorphism
algebra, which is the correct symmetry of classical gravity, naturally
involve worldlines \cite{Lar98}. This indicates that the correct
treatment of quantum gravity conceptually involves field theory, or at
least a theory with particle-like quanta.

The predictions so far were negative -- proton decay and supersymmetric 
particles are not to occur. Mainly negative predictions are
in a sense expected, since the SM is so successful experimentally, but
one would nevertheless like to have some way to discriminate between
$\mb(3|8)$ and $\ssu$. In fact, there are two generic effects that point
to $\mb(3|8)$ and that have already been experimentally confirmed: CP 
violation and several generations.

CP violation can be accounted for already within the SM by several
mechanisms (theta angle, neutrino mass matrix, CKM matrix) \cite{Ben02},
but it is one of the least understood aspects, as witnessed by expressions 
like ``strong CP problem'' and ``flavour problem'' . Also, 
CP violation in the SM is too weak to explain certain cosmological data,
in particular the observed baryon-to-photon ratio. Almost any extension
of the SM provides new sources of CP violation \cite{Nir99},
but $\mb(3|8)$ has a mechanism which makes CP violation manifest 
already at the symmetry level: conjugate representations are inequivalent.

In fact, every infinite-dimensional algebra of vector fields has this
property. E.g., in $gl(n)$ the distinction between conjugate
representations is purely conventional, because this algebra is generated
by vector fields of the form $x^\mu\dnu$, where $[\dnu, x^\mu] =
\dlt^\mu_\nu$. One usually considers $x^\mu$ as the coordinate and $\dnu
= \d/\d x^\nu$ as the derivative, but the opposite interpretation is
equally valid. In
contrast, in the full algebra $\vect(n)$ of vector fields (which is
related to $gl(n)$ as $\mb(3|8)$ is related to $\ssu$), the distinction
between co- and contra- is substantial. A general vector field is
of the form $\xi = \xmu(x) \dmu$, and thus the symmetry between $x^\mu$
and $\dnu$ is lost. On the representation level, this is
manifested in the existence of the exterior derivative, which acts
on covariant skew tensor fields only.

Let us now turn to the fermions. Because gauge bosons must be identified
with connections of the associated current algebra, they are not unified
into a single multiplet. Analogously, there is no compelling reason why
different species of quarks and leptons should be unified. Note in 
passing that all
applications of conformal field theory to statistical physics
contain several Virasoro irreps \cite{FMS96}. This is the
reason why $\mb(3|8)$, despite being a superalgebra, does not predict
unobserved SUSY partners. Note also that $\mb(3|8)$ is not technically a
SUSY, because it does not contain a Poincar\'e subalgebra.

However, any $\mb(3|8)$ irrep decomposes into several
$\ssu$ irreps under restriction, which can account for several
generations. I thus suggest ``vertical'' unification between generations,
but not ``horizontal'' unification within a single generation.
Denote the basis of a $\map(4,\mb(3|8))$ module by
$\psi(x,\th,u,\vth)$, where $x$ is the spacetime coordinate and
$(\th,u,\vth)$ are coordinates in internal $3|8$-dimensional space (see
Appendix). Upon restriction to $\map(4,\ssu)$, this module contains three
natural submodules, defined by the conditions 
$\int d^6\th\,\psi(x,\th,u,\vth)
= 0$, 
$\int d^3u\,\psi(x,\th,u,\vth) = 0$, and 
$\int
d^2\vth\,\psi(x,\th,u,\vth) = 0$, respectively. 
As a very speculative note, this might suggest that the three SM 
generations are related to depth $3$.

The predictions so far were quite generic, and apply to $\vle(3|6)$ as
well, but the fermion content is $\mb(3|8)$-specific.
To consider $\mb(3|8)$ would be overkill if we wanted to
identify fermions with tensor modules, since all information about these
is encoded already in the $\ssu$ subalgebra. Instead, I suggested in
\cite{Lar02} that fermions should be identified with closed form modules,
i.e. the irreducible quotients $\ker\nabla$, where $\nabla$ is an
invariant morphism (the analogue of the exterior derivative).
$\mb(3|8)$ tensor modules are labelled by weights
$(p,q; r; Y)$, $p,q,r\in \NN$, $Y\in\RR$, where $p\pi_1 + q\pi_2$ is an 
$su(3)$ weight, $r$ is an $su(2)$ weight and $Y$ is weak 
hypercharge\footnote{In \cite{Lar02}, the abelian weight was defined with
a factor $3$ in order to work exclusively with integers.}. 
The tensor module $T(p,q; r; Y)$ is irreducible except for four series
\cite{KR01,Lar02}\footnote{Kac and Rudakov \cite{KR01} obtain the
contragredient result.}:
\bes
\Om_A(p,r) &=& T(p,0; r; -\four p+r-2), \nl
\Om_B(p,r) &=& T(p,0; r; -\four p-r-4), \nl
\Om_C(q,r) &=& T(0,q; r; \four q+r+2), \nl
\Om_D(q,r) &=& T(0,q; r; \four q-r). 
\eens

The best assignment possible is described by the following table:
\[
\begin{array}{|c|c|c|c|c|} 
\hline
\hbox{weight} & \hbox{fermion} &\hbox{form}& Y(\hbox{SM}) & Y(\mb) \\
\hline
(0,1;1;{1\/3}) & \begin{pmatrix} u_L \\ d_L \end{pmatrix} &
\Om_D(1,1) & {1\/3} &{1\/3}
\\
(0,1;0;{4\/3}) &  u_R & \Om_D(1,0) & {4\/3} & {4\/3}
\\
(0,1;0;-{2\/3}) & d_R & \Om_C(1,0) & -{2\/3} & {10\/3}
\\
\hline
(0,0;1;-1) & \begin{pmatrix} \nu_{eL} \\ e_L \end{pmatrix} &
\Om_D(0,1) & -1 & -1
\\
(0,0;0;-2) & e_R & \Om_C(0,0) & -2 & 2
\\
\hline
\end{array}
\]
We see that the $su(3)+su(2)$ weights of the fundamental fermions can
be fitted very snugly into the list of form modules, but that the
hypercharge assignment in the C sector ($d_R$ and $e_R$) is off by four
units. Also, the anti-fermions fit into the A and B sectors, but here
we have an additional discrepancy $\Delta Y = 2$. Replacing $\mb(3|8)$
by $\vle(3|6)$ does not help; the reducibility conditions are different
\cite{KR00a,KR00b}, but it is still impossile to fit all fermions to
closed form modules.

Although the hypercharge problem is embarrassing, it can be circumvented
in several ways. Maybe $d_R$ and $e_R$ (but not the fermions in the 
D sector) should be identified with irreducible tensor modules. Another
possibility is that the restriction $\mb(3|8) \to \ssu$, which is not
understood, introduces extra hypercharge in the C sector. E.g., 
integration over the internal $3|8$-dimensional space contributes
$\Delta Y = 2$. However, we can conclude that hypercharge is quantized
in multiples of $1/3$.

To summarize, the main predictions from the assumption of maximal 
depth are:
\begin{itemize}
\item
$\ssu$ symmetry; more precisely, a localized form thereof.
\item
No proton decay, since there are no new gauge bosons.
\item
No sparticles, since supersymmetry is incompatible with maximal depth.
\item
Manifest CP violation, since conjugate representations are inequivalent.
\item
Several generations, since the restriction of an $\mb(3|8)$ irrep to
$\ssu$ is reducible.
\item
Several fermion species in qualitative agreement with experiments, but
the na\"\i ve hypercharge assignment in the C sector is off by four units.
\item
Charge quantization in units of $1/3$.
\end{itemize}

The absense of proton decay and sparticles will be tested at the LHC and
SuperKamiokande over the next decade. However, maximal depth excludes
SUSY and GUTs in principle, not only at energy scales accessible to the
next generation of experiments. The reason is simple: if nature reaches
maximal depth at some scale, it would be unnatural to retract to less
depth at higher energy. There is of course also the possibility that
proton decay and sparticles will be seen in experiments soon. If so, the
fact that maximal depth immediately leads to the correct SM symmetries 
will merely be a mathematical curiousity.

\bigskip
\noindent{\bf Appendix}
\smallskip

\noindent
Every algebra of polynomial vector fields $\oj$ admits a grading by
finite-dimensional vector spaces of depth $d$ and height $h$:
\[
\oj = \oj_{-d} + ... +\oj_{-1} + \oj_0 + \oj_1 + ... + \oj_h.
\]
Finite-dimensional Lie superalgebras have gradings with depth 
and height equal and at most $2$, whereas infinite-dimensional algebras 
have infinite height and finite depth. Apart from an inconsistent regrading 
of $\ksle(5|10)$, $\mb(3|8)$ is the unique simple Lie superalgebra of 
maximal depth $3$ \cite{Kac98,Sh99}.

The grading is said to be {\em consistent} if odd subspaces are purely 
fermionic and even subspaces purely bosonic. It is known that the only
consistently graded simple algebras are the contact algebras $\kk(1|m)$
(a.k.a. the centerless $N=m$ superconformal algebra) and the four
exceptions $\ksle(5|10)$, $\vle(3|6)$, $\mb(3|8)$ and $\kas(1|6)$.
These algebras were first discovered by Shchepochkina \cite{Sh83,Sh99}
($\kas(1|6)$ was independently found in \cite{CK97}), and their
consistent gradings first appeared in \cite{Kac98}. They were described 
explicitly in terms of generators and brackets in \cite{CK99}, and
the preserved geometries were discovered in \cite{Lar01}.

It is instructive to consider some examples.

The polynomial part of the Virasoro algebra $\vect(1)$ has generators 
$L_m = t^{m+1}\dd t$, $m \geq -1$. Setting $\deg L_m = m$, we see that 
this algebra has depth $1$; the grading is inconsistent since $L_{-1}$
is bosonic.

The polynomial part of the superconformal algebra $\kk(1|1)$ has 
generators
$L_m = t^m(t\dd t + {m+1\/2}\th\dd\th)$, $m \geq -1$, and 
$G_r = t^{r+1/2}(\dd\th + \th\dd t)$, $r\geq -1/2$.
Setting $\deg L_m = 2m$ and $\deg G_r = 2r$ makes this into a consistently
graded superalgebra of depth $2$.

The Clifford algebra $Cl(m)$ has $m$ fermionic generators $D_i$ and a
single bosonic generator $E$. The brackets read
\bes
\{D_i, D_j\} &=& 2g_{ij} E, \nl
{[}D_i, E] &=& [E,E] = 0,
\eens
where $g_{ij}$ are symmetric structure constants. $Cl(m)$ can be
realized as vector fields acting on $\CC^{1|m}$ as follows:
\bes
D_i &=& \dd{\th_i} + g_{ij}\th^j\dd t, \nl
E &=& \dd t.
\eens
The contact superalgebra $\kk(1|m) \subset \vect(1|m)$ is generated by 
vector fields $X$ which preserve the dual Pfaff equation $D_i = 0$. 
In other words, $X \in \kk(1|m)$ if
\[
[D_i, X] = f^j_i D_j,
\]
where $f^i_j$ are some functions depending on $X$. Note that a general
vector field $X\in\vect(1|m)$ satisfies
\[
[D_i, X] = f^j_i D_j + g_i E,
\]
so $\kk(1|m)$ is characterized by $g_i = 0$. Setting $\deg \th^i = 1$, 
$\deg t = 2$ gives $\kk(1|m)$ a consistent grading of depth $2$. 
$\kk(1|m)$ is known in physics as the centerless $N=m$ superconformal 
algebra, which is well known to have a central extension iff 
$m\leq4$ \cite{GLS97}. In general, a Virasoro-like extension can be 
obtained by restriction from $\vect(1|m)$;
this cocycle is evidently non-central for $m\geq5$.

Let us now turn to the definition of $\mb(3|8)$ \cite{Lar01}.
Consider the following nilpotent Lie superalgebra $\toj_-$:
\bes
\{D^{ia}, D^{jb}\} &=& -6\eps^{ijk}\eps^{ab}E_k, \nl 
{[}D^{ia}, E_j] &=& 2\dlt^i_j \eps^{ab} F_b, \nl
\{D^{ia}, F_b\} &=& [E_i, E_j] = [E_i, F_a] = \{F_a, F_b\} = 0,
\eens
where $i,j,k=1,2,3$ are $\CC^3$ indices, $a,b=1,2$ are $\CC^2$ indices,
and $\eps^{ijk}$ and $\eps^{ab}$ are the totally anti-symmetric constant
tensors. Setting $\deg D^{ia} = -1$, $\deg E_i = -2$, $\deg F_a = -3$ 
makes $\toj_-$ into a graded superalgebra; the grading is evidently
consistent.
Verification of the Jacobi identities is non-trivial; one needs that the 
three generators $F_a$, $F_b$ and $F_c$ can never be linearly independent.

Consider $\CC^{3|8}$ with basis spanned by three even coordinates
$u^i$, six odd coordinates $\th_{ia}$, and two
more odd coordinates $\vth^a$,
where $\deg \th_{ia} = 1$, $\deg u^i = 2$ and $\deg \vth^a = 3$.
$\toj_-$ can be realized as a subalgebra of $vect(3|8)$ as follows:
\bes
D^{ia} &=& \dd{\th_{ia}} - 3\eps^{ijk}\th^a_j\dd{u_k}
+ \eps^{ijk}\th^a_j\th^b_k\dd{\vth^b} - \eps^{ab}u^i\dd{\vth^b}, \nl
E_i &=& \dd{u^i} - \th^a_i\dd{\vth^a}, \nl
F_a &=& \dd{\vth^a},
\eens
where $\th^a_i = \eps^{ab}\th_{bi}$.
$\toj_-$ is naturally an $\ssu$ module;
this is where contact with the SM symmetries is made.
$su(3)$ acts on $\CC^3$, $su(2)$ acts on $\CC^2$ and $u(1)$ computes
the grading; the grading operator is
\[
Z = 3Y = 3\vth^a\dd{\vth^a} + 2u^i\dd{u_i} + \th_{ia}\dd{\th_{ia}},
\]
where $Y$ is weak hypercharge.

$\mb(3|8)$ is now defined as the subsuperalgebra of $\vect(3|8)$ which 
preserves the dual Pfaff equation $D^{ia}=0$.
In other words, $X\in\mb(3|8)$ iff 
\[
[D^{ia}, X] = f^{ia}_{jb}D^{jb},
\]
for some functions $f^{ia}_{jb}$ (depending on $X$). 
An equivalent definition $\mb(3|8)$ is as the Cartan prolong 
$(\oj_-, \oj_0)_*$, where $\oj_-$ is isomorphic to $\toj_-$ but 
commutes with it. This means that we start from the realizations of
$\oj_-$ and $\oj_0$ as vector fields on $\CC^{3|8}$, and define
$\oj_k$, $k > 0$, recursively as the maximal subalgebra of $\vect(3|8)$
such that $[\oj_{-1}, \oj_k] \subset \oj_{k-1}$.

$\vle(3|6)$ is defined in a similar manner, except that we set $F_a = 0$
in $\toj_-$. Since the realization requires
two less fermionic coordinates, this is a subalgebra of $\vect(3|6)$.

\end{document}